\documentclass[aip,apl,12pt,preprint]{revtex4-1}
\usepackage{amsmath,amssymb,graphicx}
\draft 

\begin{document}


\title{Dimensional Bounds on Vircator Emission} 



\author{J. I. Katz}
\email[]{katz@wuphys.wustl.edu}
\affiliation{Dept. of Physics and McDonnell Center for the Space Sciences\\
1 Brookings Dr. CB1105 Washington University\\St.~Louis, Mo. 63130}
\affiliation{MITRE Corp. 7525 Colshire Dr., McLean, Va. 22102}


\date{\today}

\begin{abstract}
Vircators (Virtual Cathode Oscillators) are sources of short-pulsed, high
power, microwave (GHz) radiation.  An essentially dimensional argument
relates their radiated power, pulse energy and oscillation frequency to
their driving voltage and fundamental physical constants.  For a diode
of width and gap 10 cm and for voltages of a few hundred keV the peak
radiated power cannot exceed ${\cal O} (\text{30 GW})$ and the broad-band
single cycle radiated energy cannot exceed ${\cal O}(\text{3 J})$.  If
electrons can be accelerated to relativistic energies higher powers and
radiated energies may be possible.
\end{abstract}


\maketitle 

Vircators (Virtual Cathode Oscillators) are sources of short-pulsed,
high power, microwave radiation at multi-GHz frequencies.  In a Vircator
a sharply rising voltage pulse produces an intense pulse of field-emitted
electrons.  The electrons are attracted by a mesh anode.  Most of them pass
through the gaps in the anode, forming a ``virtual cathode''.  If the
injected current density exceeds its space charge limit, a portion of the
electron cloud reverses its motion as a result of its self-repulsion and the
attractive force of the anode.  The resulting oscillation radiates a short
but intense pulse of microwaves.  Maximizing net charge of the cloud and its
acceleration maximizes the radiated power.  Although there is an extensive
literature \cite{M77,M82,S83,T87,BS92,NABG94,GN97,BS01,BSS07,B09,DV09} and a
wide variety of designs have been developed, comparatively little attention
has been paid to fundamental limits on vircator performance and its scaling
with driving voltage.  These limits

Here we demonstrate upper limits to the radiated power of a vircator in
a simple, essentially order-of-magnitude, model.  Real vircators are more
complicated---their radiation propagates in a waveguide rather than in
free space as we assume, their electrons have a complicated distribution
in phase space that must be calculated numerically, rather than our crude
model described by a single length, density and velocity, the potential
is also a complicated, numerically calculated, function, rather than a
single static scalar as we assume, radiation by semi-relativistic electrons
should be calculated numerically rather than from the result for dipole
radiation (in the nonrelativistic limit) that we use.

Our model for radiated power has only one free parameter, the driving
potential, and therefore leads to a simple result that is likely to bound
the power produced in real, more complex, vircators.  Our bound on the
radiated energy also includes the size scale of the diode as a parameter.
It is not possible to derive a single bound for more complete models and
calculations of charges and currents in vacuum diodes \cite{S05,SST07}
(these authors did not calculate the emitted radiation), because they must
be described by several independent parameters.  A quantitative bound would
require optimization over a several-dimensional parameter space; its origin
would not be apparent and its numerical value would depend on detailed
geometric assumptions.  However, it would scale as our bounds do; only the
numerical coefficients would be different.  The origin and scaling of our
simple bounds are transparent.

In the nonrelativistic limit our model consists of the following relations:
\begin{align}
d &= Qr\\
Q &= n r^3 e\\
V &= {Q \over r}\\
\alpha &\equiv {eV \over m_e c^2}\\
\omega = \omega_p &\equiv \sqrt{4 \pi n e^2 \over m_e}.
\end{align}
Here $r$ is a length scale, $Q$ is the charge of the electron cloud, $n$
is its number density, $V$ is a characteristic electrostatic potential, $d$
is the magnitude of the oscillating dipole moment, $\omega_p$ is the nominal
electron plasma frequency, which approximates the characteristic frequency
$\omega$ of oscillation of the electron cloud and $\alpha$ is a
dimensionless parameter describing the characteristic potential, and is
of the same order as the imposed potential between cathode and anode.  The
remaining variables are the fundamental physical constants $e$, $m_e$ and
$c$.

We assumed that there is only one length scale $r$, that may be taken
as the distance between cathode and anode.  If the physical cathode is
smaller, the electron cloud broadens to approximately this width.  If the
physical cathode and anode have areas $A \gg r^2$, as in a parallel-plate
capacitor, the problem is essentially that of $A/r^2$ vircators in parallel,
and our results should be multiplied by this ratio.

Combining these relations and using the relation \cite{J75}
\begin{equation}
P = {1 \over 3}{d^2 \omega^4 \over c^3}
\end{equation}
for the power $P$ radiated by an oscillating electric dipole in free space,
we find
\begin{equation}
\label{Pnr}
P = {(4\pi)^2 \over 3} \alpha^4 {m_e^2 c^5 \over e^2} = {(4\pi)^2 \over 3}
\alpha^4\ \text{8.8 GW}.
\end{equation}
The numerical coefficient is 3.3 for $\alpha = 0.5$, representative of
vercators in practice.  The expression for electric dipole radiation is
applicable only for $\alpha \ll 1$, and is expected to be wrong for $\alpha
> 0.5$

The electrons' oscillations rapidly dephase because of self-shielding that
varies within the electron cloud \cite{BB63}.  As a result, the
characteristic width of the pulse of radiation emitted by an instantaneous
voltage pulse is $\sim 1/\omega$.  The radiated energy
\begin{equation}
\label{Enr}
{\cal E} = {1 \over 3} {d^2 \omega^3 \over c^3} = {(4\pi)^{3/2} \over 3}
\alpha^{7/2} (m_e c^2) \left({r m_e c^2 \over e^2}\right) = {(4\pi)^{3/2}
\over 3} \alpha^{7/2} \left({r \over \text{10 cm}}\right)\ \text{3 J}.
\end{equation}
Although the instantaneous power is high, the radiated energy is small.
The spectrum has the characteristic frequency
\begin{equation}
\nu = \nu_p = \sqrt{4 \pi \alpha}{c \over r} = \sqrt{\alpha \over \pi}
\left({\text{1 cm} \over r}\right) 30\ \text{GHz}.
\end{equation}
These numerical values are consistent with measurements in the
nonrelativistic regime \cite{S83}.

Comparing the energy (\ref{Enr}) to the electrostatic energy $Q^2/r$ of the
electron cloud leads to a radiation efficiency
\begin{equation}
\epsilon = {(4\pi)^{3/2} \over 3} \alpha^{3/2},
\end{equation}
valid only in the limit $\alpha \ll 1$.  This is greater than measured
\cite{S83} efficiencies of vircator radiation for $\alpha \sim 0.5$, that
are a few percent or less.  This may be attributed to use of a rough
approximation of the actual charge distribution and motion and to neglect
of radiation reaction, which damps the motion of radiating charges, but
which would involve a number of well-known paradoxes \cite{J75}. 

Analogous estimates are possible in the ultra-relativistic limit.  The
relativistic form of the Larmor expression for the radiation by an 
accelerated charge, with acceleration parallel to the velocity (the same
electric field $E$ gives the electron cloud its relativistic velocity and
further accelerates it) is
\begin{equation}
P = {2 \over 3}{Q^2 a^2 \gamma^6 \over c^3},
\end{equation}
where $\vec a$ is the acceleration, and $\gamma$ the Lorentz factor.  For
${\vec a} \parallel {\vec v}$
\begin{equation}
a = {e E \over m_e \gamma^3},
\end{equation}
so the nonrelativistic result
\begin{equation}
P = {2 \over 3} {Q^2 e^2 \over c^3 m_e^2} E^2
\end{equation}
is recovered.

Substituting $Q = \alpha m_e c^2 r/e$ and $E = \alpha m_e c^2/(er)$ yields
\begin{equation}
P = {2 \over 3} \alpha^4 {m_e^2 c^5 \over e^2}.
\end{equation}
Aside from the numerical factor, this is the same result as Eq.~\ref{Pnr}
obtained in the nonrelativistic limit.  The characteristic time scale of
emission is now $r/c$, rather than $1/\omega_p$, and the corresponding
radiated energy is
\begin{equation}
E = P {r \over c} = {2 \over 3} \alpha^4 {m_e^2 c^4 \over e^2} r.
\end{equation}
The scaling with voltage is slightly different than that of the
nonrelativistic result Eq.~\ref{Enr}, but the dimensional factor is the
same.  The implied efficiency would be
\begin{equation}
\epsilon = {2 \over 3} \alpha^2.
\end{equation}
This would lead to the impossible result $\epsilon > 1$ in the
ultra-relativistic limit $\alpha \gg 1$, a discrepancy that is explained
by the neglect of radiation reaction.

The fact that the dimensional factors in the results for $P$ and $E$
are the same in the nonrelativistic and relativistic limits is unavoidable.
There are only three dimensional quantities in the problem, once the
electrostatic potential has been scaled to the electron rest mass, and
therefore only one possible form for power and energy.

These are general limits on the performance of vircators, expressed in
terms of fundamental physical constants, that are insensitive to details of
design.  They are consistent with measured performance; vircators
readily emit powers of a few GW with driving potentials of a few hundred
kV ($\alpha \approx 0.5$), but have not emitted orders of magnitude more.
Although the emitted power is a steeply increasing function (proportional
to the fourth power) of applied voltage in both nonrelativistic and
ultra-relativistic regimes, the voltage than can be applied is limited by
both available low-impedance power supplies (the voltage must rise rapidly
in order than field emission not short out the potential before the full
voltage is applied) and by the requirement that the voltage not be
short-circuited by breakdown outside the diode.

I thank R. L. Garwin and K. Pister for useful discussions.


%
%

%


\bibliography{vircator.bib}

\end{document}